\pdfoutput=1
\documentclass[a4paper,aip,american,citeautoscript,floatfix,jcp,pdftex,superscriptaddress,%
longbibliography,%
reprint,twocolumn%
]{revtex4-1}

\usepackage{amsmath,amssymb}
\usepackage{color}
\usepackage[T1]{fontenc}
\usepackage{graphicx}
\usepackage[utf8]{inputenc}
\usepackage{microtype}
\usepackage{newpxtext,newpxmath}
\usepackage{textcomp}
\usepackage[textsize=footnotesize]{todonotes}
\usepackage{xspace}
\usepackage{hyperref, hypernat}
\usepackage[ams,todos]{CMImacros}

\newlength{\figwidth}
\newcommand*{\NHHH}{\ensuremath{\text{NH}_3}\xspace}%
\newcommand*{\nNHHH}{\ensuremath{{}^{14}\text{NH}_3}\xspace}%

\newcommand{\cfeldesy}{\affiliation{Center for Free-Electron Laser Science, Deutsches
      Elektronen-Synchrotron DESY, Notkestrasse 85, 22607 Hamburg, Germany}}%
\newcommand{\uhhcui}{\affiliation{The Hamburg Center for Ultrafast Imaging, Universität Hamburg,
      Luruper Chaussee 149, 22761 Hamburg, Germany}}%
\newcommand{\uhhphys}{\affiliation{Department of Physics, Universität Hamburg, Luruper Chaussee 149,
      22761 Hamburg, Germany}}%
\newcommand{\cmiweb}{\homepage{https://www.controlled-molecule-imaging.org}}%
\newcommand{\ayemail}{\email{andrey.yachmenev@cfel.de}}%

\begin{document}
\title{General variational approach to nuclear-quadrupole coupling in rovibrational spectra of
   polyatomic molecules}%
\author{Andrey Yachmenev}\ayemail\cmiweb\cfeldesy\uhhcui%
\author{Jochen Küpper}\cfeldesy\uhhcui\uhhphys%
\date{\today}%
\begin{abstract}
   A general algorithm for computing the quadrupole-hyperfine effects in the rovibrational spectra
   of polyatomic molecules is presented for the case of ammonia (\NHHH). The method extends the
   general variational approach TROVE by adding the extra term in the Hamiltonian that describes the
   nuclear quadrupole coupling, with no inherent limitation on the number of quadrupolar nuclei in a
   molecule. We applied the new approach to compute the nitrogen-nuclear-quadrupole hyperfine
   structure in the rovibrational spectrum of \nNHHH. These results agree very well with recent
   experimental spectroscopic data for the pure rotational transitions in the ground vibrational and
   $\nu_2$ states, and the rovibrational transitions in the $\nu_1$, $\nu_3$, $2\nu_4$, and
   $\nu_1+\nu_3$ bands. The computed hyperfine-resolved rovibrational spectrum of ammonia will be
   beneficial for the assignment of experimental rovibrational spectra, further detection of ammonia
   in interstellar space, and studies of the proton-to-electron mass variation.
\end{abstract}
\maketitle

Precision spectroscopy of small molecules promises new windows into fundamental
physics~\cite{DeMille:PT68:34}, but requires a detailed understanding of their complex energy-level
structures, including the hyperfine structure due to the interaction between the rovibronic
molecular states and the nuclear spins of their constituent atoms. Recent advances in
high-resolution sub-Doppler spectroscopy and astronomical observations of interstellar environments
have also triggered renewed interest in the hyperfine-structure of molecular spectra. For instance,
the hyperfine structure of molecular rovibrational energy levels is relevant for a variety of
applications in precision spectroscopy~\cite{Bethlem:EPJST163:55, Schnell:FD150:33}, ultracold
chemical reactions~\cite{Bell:MP107:99, Meerakker:CR112:4828, Naulin:IRPC33:427, Stuhl:ARPC65:501},
highly correlated quantum gases~\cite{Baranov:ChemRev112:5012, Ospelkaus:PRL104:030402,
   Aldegunde:PRA80:043410, Moses:NatPhys13:13}, and quantum-information
processing~\cite{Wei:JCP135:154102, Jaouadi:JCP139:014310}. Shifts of the hyperfine-split energy
levels and transition frequencies in time can serve as useful probe for testing whether there is any
cosmological variability of the proton-to-electron mass ratio~\cite{Veldhoven:EPJD31:337,
   Flambaum:PRL98:240801, Owens:PRA93:052506, Cheng:PRL117:253201}, predicted by theories beyond the
Standard Model. Even for spectra with Doppler-limited resolution, detailed knowledge of the
hyperfine patterns of rovibrational energy levels is very useful for determining the absolute line
positions and verifying the spectroscopic assignments~\cite{Twagirayezu:JCP145:144302}.

So far the hyperfine structure of rovibrational energy levels is described using effective
Hamiltonian models~\cite{Hougen:JCP57:4207, Gordy:MWMolSpec}, with only a few
exceptions~\cite{Jensen:JMolSpec150:137, Jensen:MolPhys91:319, Miani:JCP120:2732}. However, due to
the scantness of hyperfine-resolved spectroscopic data, the effective Hamiltonian approaches are
very limited in extrapolating to energy levels that are not directly experimentally sampled. For
accurate predictions over a broader spectral range it is highly desirable to employ variational
approaches, which show much better extrapolation properties than the effective-Hamiltonian
approaches, \eg, because they intrinsically incorporate all resonant interactions between
rovibrational states. Over the last decade, a considerable amount of work has been put into the
development of such variational approaches and their efficient computer
implementations~\cite{Lauvergnat:JCP116:8560,Yurchenko:JMS245:126, Matyus:JCP127:084102,
   Matyus:JCP130:134112, Wang:JCP130:094101, Avila:JCP139:134114, Fabri:JCP134:074105,
   Yachmenev:JCP143:014105}. TROVE~\cite{Yurchenko:JMS245:126, Yachmenev:JCP143:014105} is a general
black-box computational paradigm for calculating the rovibrational energy levels of polyatomic
molecules in isolated electronic states. It is based on a completely numerical approach for
computing the Hamiltonian matrix and solving the eigenvalue problem. Along with its algorithmic
efficiency, TROVE benefits from the use of molecular symmetry, including non-Abelian symmetry
groups~\cite{Yurchenko:JCTC:submitted}, curvilinear internal coordinates and the Eckart coordinate
frame~\cite{Yachmenev:JCP143:014105}. Over the last few years TROVE has actively been employed for
computing comprehensive rovibrational line lists for a number of polyatomic molecules important for
modeling and characterisation of cool stars and exoplanets~\cite{Tennyson:JMS327:73}.

Here, we extend TROVE to include hyperfine effects at the level of the nuclear-quadrupole coupling.
The coupling is described by the interaction of the nuclear quadrupole moments with the electric
field gradient at the nuclei. The latter is treated as a function of the nuclear internal
coordinates and we impose no inherent limitations on the number of internal degrees of freedom nor
the number of quadrupolar nuclei. Thus, it is applicable to molecules with arbitrary structure. To
our knowledge, this work is the first attempt to create a general approach for computing nuclear
spin effects with high accuracy based on the robust variational method.

We apply the newly developed method to compute the rovibrational spectrum of \NHHH with resolved
hyperfine quadrupole structure. We use existing accurate potential energy and electric dipole moment
surfaces of \NHHH~\cite{Yurchenko:JMolSpec268:123, Yurchenko:JPCA113:11845} along with a newly
\emph{ab initio} calculated electric field gradient tensor surface. Small empirical corrections were
added to the calculated spin-free vibrational band centers to match the experimentally derived
energies from the MARVEL database~\cite{Derzi:JQSRT161:117}. The resulting calculated data are in
very good agreement with the available experimental quadrupole splittings for the ground vibrational
state~\cite{Coudert:AA449:855} and the excited vibrational states
$\nu_2$~\cite{Belov:JMolSpec189:1}, $\nu_1$, $\nu_3$, and $2\nu_4$~\cite{Dietiker:JCP143:244305} as
well as with the hyperfine splittings and intensities of recent sub-Doppler spectral measurements
for the $\nu_1+\nu_3$ band~\cite{Twagirayezu:JCP145:144302}. We expect that our results will aid the
spectroscopic analysis of many unassigned overlapping rovibrational features in the spectrum of
\nNHHH and support astronomical detection of ammonia in interstellar media. The results are also
relevant for future laboratory and astronomical observations of the proton-to-electron mass
variation~\cite{Veldhoven:EPJD31:337, Owens:PRA93:052506}.


Within the Born-Oppenheimer approximation the quadrupole structure of the rovibrational energy
levels in a molecule containing $l=1\ldots{}N$ quadrupolar nuclei is described by the coupling of
the electric field gradient (EFG) at each $l$\/th nucleus $\mathbf{V}(l)$ with its quadrupole moment
$\mathbf{Q}(l)$~\cite{Cook:AJP39:1433}
\begin{eqnarray}
  H_\text{qc} = \sum_l^N \mathbf{V}(l) \cdot \mathbf{Q}(l).
\end{eqnarray}
The operator $\mathbf{V}(l)$ acts only on the rovibrational coordinates and momenta of the nuclei,
while the operator $\mathbf{Q}(l)$ depends solely on the nuclear spin angular momenta and
fine-structure constants. The total spin-rovibrational wave functions \ket{F,m_F} can be constructed
from the linear combinations of products of the rovibrational wave functions \ket{J,m_J,w} and the
nuclear spin functions \ket{I,m_I,\mathcal{I}}. Here, $J$, $I$, and $F$ denote the quantum numbers
of the rotational $\hat{\op{J}}$, collective nuclear spin $\hat{\op{I}}$, and total angular momentum
$\hat{\op{F}}=\hat{\op{J}}+\hat{\op{I}}$ operators, respectively. $m_J$, $m_I$, and $m_F$ denote the
quantum numbers of the corresponding laboratory-frame projections, and $w$ and $\mathcal{I}$ stand
for the sets of additional quanta characterizing rovibrational and nuclear spin states,
respectively. The matrix representation of the quadrupole-coupling Hamiltonian in a basis of
functions \ket{F,m_F} is diagonal in $F$ and $m_F$ and can be expressed as
\begin{multline}
   \braopket{F,m_F}{H_\text{qc}}{F,m_F}%
   = (-1)^{J+I'+F} \left\{\begin{array}{ccc} F & I' & J' \\ 2 & J & I\end{array}\right\} \\
   \times\sum_l^N\branormket{J',w'}{V^{(2)}(l)}{J,w}\cdot\branormket{I',\mathcal{I}'}{Q^{(2)}(l)}{I,\mathcal{I}},
   \label{eq:hqc}
\end{multline}
where $V^{(2)}(l)$ and $Q^{(2)}(l)$ are the EFG and the quadrupole moment operators in the
irreducible tensor form.

For $N>1$ quadrupolar nuclei with spins $I_1,I_2,\ldots{}I_N$, the nuclear spin basis functions
$|I,m_I,{\mathcal I}\rangle$ are unambiguously characterized by the set of quantum numbers
${\mathcal I}=\{ I_{12},I_{13},...,I_{1N-1}\}$ and $I\equiv I_{1N}$ denoting the eigenstates of the
coupled spin angular momenta $\hat{\bf I}_{12} = \hat{\bf I}_1+\hat{\bf I}_2$,
$\hat{\bf I}_{13} = \hat{\bf I}_{12}+\hat{\bf I}_3$, ...,
$\hat{\bf I}_{1N-1} = \hat{\bf I}_{1N-2}+\hat{\bf I}_{N-1}$ and
$\hat{\bf I}_{1N} = \hat{\bf I}_{1N-1}+\hat{\bf I}_N$, respectively. For $N=1$, the quantum number
${\mathcal I}$ is omitted. In this basis, the reduced matrix elements of the quadrupole moment
operator $Q^{(2)}(l)$ can be expressed as
\begin{equation}
   \label{eq:qmom_red}
   \branormket{I',\mathcal{I}'}{Q^{(2)}(l)}{I,\mathcal{I}}%
   = \frac{1}{2}(eQ_l)C_{l}^{(I'\mathcal{I}',I\mathcal{I})}
   \left(\begin{array}{ccc} I_l & 2 & I_l \\ -I_l & 0 & I_l\end{array}\right)^{-1}
\end{equation}
where $(eQ_l)$ stands for the nuclear quadrupole constant and the explicit expression for the
coupling coefficients $C_{l}^{(I'{\mathcal I}',I{\mathcal I})}$ is given in the supplementary
material \cite{Yachmenev:nh3_nucquad_suppinfo}.

The reduced matrix elements of the EFG operator $V^{(2)}(l)$ are given by the expression
\begin{multline}
  \label{eq:vpot_red}
  \branormket{J',w'}{V^{(2)}(l)}{J,w} = \frac{1}{2} (-1)^{J-J'}%
  \left(\begin{array}{ccc} J' & 2 & J \\ -J & 0 & J\end{array}\right)^{-1} \\
  \times\braopket{J',m'_J=J,w'}{V_{ZZ}(l)}{J,m_J=J,w},
\end{multline}
where $V_{ZZ}$ is the $(Z,Z)$-component of the EFG tensor in the laboratory frame and
$m'_J=m_J=\min(J,J')$.

The rovibrational wave functions \ket{J,m_J,w} and energies $E_{J,w}$ are obtained from TROVE
calculations and are represented by linear combinations of pure vibrational \ket{v} and symmetric
top \ket{J,m_J,k,\tau} wavefunctions
\begin{eqnarray}
  \ket{J,m_J,w} = \sum_{k\tau v} C_{k\tau v}^{(J,w)} \ket{v} \ket{J,m_J,k,\tau} ,
  \label{eq:rovibbas}
\end{eqnarray}
where $k=0\ldots{}J$; $\tau=0$ or $1$ defines rotational parity as $(-1)^\tau$, and
$C_{k\tau{}v}^{(J,w)}$ are the eigenvector coefficients of the total rovibrational Hamiltonian. To
compute the matrix elements of $V_{ZZ}$ in \eqref{eq:vpot_red} in the basis of the rovibrational
functions given by \eqref{eq:rovibbas}, we employ a general approach~\cite{Yachmenev:JCP:inprep}:
\begin{equation}
   \label{eq:vpot}
   \braopket{J',m_J',w'}{V_{ZZ}(l)}{J,m_J,w} = \mathcal{M}_{2}^{(J'm_J',Jm_J)} \mathcal{K}_{2}^{(J'w',Jw)}(l),
\end{equation}
with
\begin{multline}
   \label{eq:mmat}
   \mathcal{M}_{2}^{(J'm_J',Jm_J)} = (-1)^{m_J'}\sqrt{(2J'+1)(2J+1)} \\
   \times \sum_{\sigma=-2}^2 \left[T^{(2)}\right]^{-1}_{ZZ,\sigma}%
   \left(\begin{array}{ccc} J & 2 & J' \\ m_J & \sigma & -m_J'\end{array}\right)
\end{multline}
and
\begin{multline}
   \label{eq:kmat}
   \mathcal{K}_{2}^{(J'w',Jw)}(l) = \sum_{\substack{k'\tau' v'\\ k\tau v}}%
   \left[C_{k'\tau' v'}^{(J',w')}\right]^* \, C_{k\tau v}^{(J,w)}
   \sum_{\pm k',\pm k} \left[c_{k'}^{(\tau')}\right]^*c_{k}^{(\tau)} \\
   \times (-1)^{k'} \sum_{\sigma=-2}^2\sum_{\alpha\leq\beta}%
   \left(\begin{array}{ccc} J & 2 & J' \\ k & \sigma & -k'\end{array}\right)
   T^{(2)}_{\sigma,\alpha\beta} \braopket{v'}{\bar{V}_{\alpha\beta}(l)}{v}.
\end{multline}
Here, $\bar{V}_{\alpha\beta}(l)$ ($\alpha,\beta=x,y,z$) denotes the EFG tensor in the molecular
frame, the constant $5\times6$ matrix $\op{T}^{(2)}$ and its inverse are the forward and backward
transformation, respectively, of the traceless second-rank Cartesian tensor operator into its
spherical tensor representation, and $c_{\pm{}k}^{(\tau)}$ are the so-called Wang transformation
coefficients defining
$\ket{J,m_J,k,\tau}=c_{+k}^{(\tau)}\ket{J,m_J,k}+c_{-k}^{(\tau)}\ket{J,m_J,-k}$. Explicit
expressions for $\op{T}^{(2)}$ and $c_{\pm k}^{(\tau)}$ are given in the supplementary material \cite{Yachmenev:nh3_nucquad_suppinfo}.

The computational procedure can be summarized as following: First we variationally solve the
spin-free rovibrational problem with TROVE and obtain the rovibraitonal energies $E_{J,w}$ and wave
functions \ket{J,m_J,w} according to \eqref{eq:rovibbas} for states with $J=0\ldots{}J_\text{max}$
and energies $E_{J,w}$ below a given threshold. The \emph{ab initio} computed values of the EFG
tensor $\bar{V}_{\alpha\beta}(l)$ at various molecular geometries are obtained from a least-squares
fit by a truncated power series expansions in terms of the internal coordinates of the molecule. In
the next step, the matrix elements $\mathcal{K}_2(l)$ in \eqref{eq:kmat} are evaluated for the
desired rovibrational states for all quadrupolar nuclei $l$ and stored in a database format similar
to that adopted for spectroscopic line lists~\cite{Tennyson:JMS327:73}. The matrix elements of
$H_\text{qc}$ in \eqref{eq:hqc} can efficiently be assembled on the fly from the products of the
compact $\mathcal{K}_2(l)$ and $\mathcal{M}_2$ matrices together with the reduced matrix elements of
the quadrupole moment operator, given by \eqref{eq:qmom_red}. The spin-rovibrational energies and
wave functions are obtained by solving the eigenvalue problem for the total Hamiltonian, which is
the sum of the diagonal representation of the pure rovibrational Hamiltonian
$E_{J,w}\delta_{\mathcal{I'},\mathcal{I}}\delta_{I',I}$ and the non-diagonal matrix representation
of $H_\text{qc}$. Since the $H_\text{qc}$ operator commutes with all operations of the molecular
symmetry group, it factorizes into independent blocks for each symmetry, which are processed
separately. The relevant equations for the dipole transition line strengths and notes on the
computer implementation are presented in the supplementary material \cite{Yachmenev:nh3_nucquad_suppinfo}.


Now, we apply the developed method to compute the hyperfine quadrupole structure in the
rovibrational spectrum of \nNHHH. To compute the spin-free rovibrational states we used the approach
described in a previous study that generated an extensive rovibrational line list of
\NHHH~\cite{Yurchenko:MNRAS413:1828}. We used the available spectroscopically refined potential
energy surface (PES)~\cite{Yurchenko:JMolSpec268:123} and \emph{ab initio} dipole moment
surface~\cite{Yurchenko:JPCA113:11845} of \NHHH and truncated the vibrational basis set at the
polyad number $P=14$.

The EFG tensor $\bar{V}_{\alpha\beta}$ at the quadrupolar nucleus $^{14}$N was computed on a grid of
4700 different symmetry-independent molecular geometries of \NHHH employing the CCSD(T) level of
theory in the frozen-core approximation and the aug-cc-pVQZ basis set~\cite{Dunning:JCP90:1007,
   Kendall:JCP96:6796}. Calculations were performed using analytical coupled cluster energy
derivatives~\cite{Scuseria:JCP94:442}, as implemented in the CFOUR program package
\cite{CFOUR:2017}. To represent each element of the $\bar{V}_{\alpha\beta}$ tensor analytically, in
terms of internal coordinates of \NHHH, we first transformed it into a symmetry-adapted form under
the $\op{D}_{3h}$(M) molecular symmetry group, as
\begin{align}
  \label{eq:v_mb1} V_1^{(A_1')}  &= V_{44}, \\
  \label{eq:v_mb2} V_2^{(E_a')}  &= (2V_{12} - V_{13} - V_{23})/\sqrt{6}, \\
  \label{eq:v_mb3} V_2^{(E_b')}  &= (V_{13} - V_{23})/\sqrt{2}, \\
  \label{eq:v_mb4} V_3^{(E_a'')} &= (2V_{14} - V_{24} - V_{34})/\sqrt{6}, \\
  \label{eq:v_mb5} V_3^{(E_b'')} &= (V_{24} - V_{34})/\sqrt{2}.
\end{align}
Since the EFG tensor is traceless, the totally symmetric combination
$V^{(A_1')}=(V_{11}+V_{22}+V_{33})/\sqrt{3}=0$ vanishes. $V_{ij}$ ($i,j=1\ldots{}4$) denote
projections of the $\bar{V}_{\alpha\beta}$ Cartesian tensor onto a system of four molecular-bond
unit vectors, defined as following~\cite{Yurchenko:JPCA113:11845}
\begin{align}
  \label{eq:mb_system}
  \op{e}_i &= (\op{r}_{\text{H}_i}-\op{r}_\text{N})/r_{\text{NH}_i} \quad \text{for~} i=1,2,3 \\
  \op{e}_4 &= \frac{\op{e}_1\times\op{e}_2 + \op{e}_2\times\op{e}_3 + \op{e}_3\times\op{e}_1}{%
             \|\op{e}_1\times\op{e}_2 + \op{e}_2\times\op{e}_3 + \op{e}_3\times\op{e}_1\|},
\end{align}
where $\op{r}_{\text{H}_i}$ and $\op{r}_\text{N}$ are the instantaneous Cartesian coordinates of the
hydrogen and nitrogen nuclei and $r_{\text{NH}_i}$ are the N--H$_i$ internuclear distances. The
$E_a$ and $E_b$ symmetry components of the doubly degenerate representations $E'$ and $E''$ are
connected by a simple orthogonal transformation and can be parametrized by one set of constants. The
remaining three symmetry-unique combinations $V_1^{(A_1')}$, $V_2^{(E_a')}$, and $V_3^{(E_a'')}$
were parametrized by the symmetry-adapted power series expansions to sixth order using the least
squares fitting. The values of the optimized parameters and the Fortran~90 functions for computing
$V_i^{(\Gamma)}$ in \eqref{eq:v_mb1}--\eqref{eq:v_mb5} are provided in the supplementary material \cite{Yachmenev:nh3_nucquad_suppinfo}.

The computed rovibrational line list for \nNHHH covers all states with $F\leq15$
($F=\abs{J-I_\text{N}}\ldots{}J+I_\text{N}$ and $I_\text{N}=1$) and energies
$E_{J,w}\leq8000~\invcm$ relative to the zero-point level. A value of $eQ=20.44$~mb for the $^{14}$N
nuclear quadrupole constant was used~\cite{Pyykko:MolPhys106:1965}.


\begin{figure}
   \includegraphics[width=\linewidth]{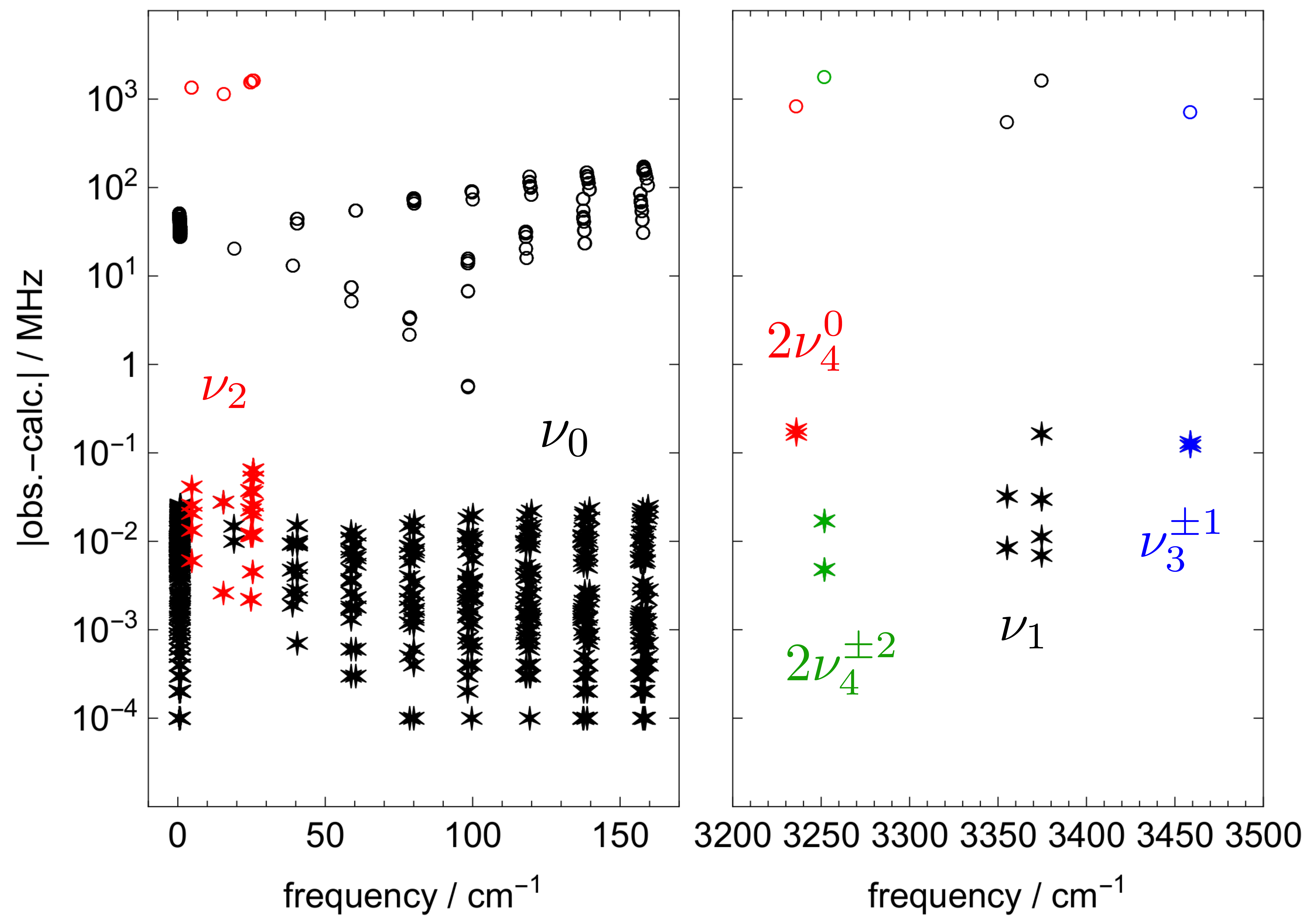}
   \caption{Absolute values of the discrepancies of the calculated transition frequencies of \NHHH
      to the experimental data for (left panel) the ground $\nu_0$ and $\nu_2$ vibrational states
      and (right panel) the $\nu_1$, $\nu_3^{\pm1}$, $2\nu_4^0$, $2\nu_4^{\pm2}$ bands. The errors
      in the rovibrational frequencies are plotted with circles while the relative errors of the
      quadrupole splittings are plotted with stars.}
   \label{fig:1}
\end{figure}
In \autoref{fig:1} we compare the predicted quadrupole hyperfine transition frequencies for \NHHH
with the available experimental data. We have chosen the most recent and easily digitized
experimental data sets, which contain rotational transitions in the ground
vibrational~\cite{Coudert:AA449:855} and $\nu_2$~\cite{Belov:JMolSpec189:1} states, and
rovibrational transitions from the ground to the $\nu_1$, $\nu_3^{\pm1}$, $2\nu_4^0$, and
$2\nu_4^{\pm2}$ vibrational states~\cite{Dietiker:JCP143:244305}. An extended survey of the
published experimental and theoretical data for the quadrupole hyperfine structure of ammonia can be
found elsewhere~\cite{Dietiker:JCP143:244305, Augustoviov:APJ824:147}. The absolute errors in the
rovibrational frequencies, plotted with circles, are within the accuracy of the underlying
PES~\cite{Yurchenko:JMolSpec268:123}. To estimate the accuracy of the predicted quadrupole
splittings, we subtracted the respective error in the rovibrational frequency unperturbed from the
quadrupole interaction effect for each transition. The resulting errors, plotted with stars in
\autoref{fig:1}, range from 0.1 to 25~kHz for the ground vibrational state and from 2.6 to 64~kHz
for the $\nu_2$ state. These values correspond to the maximal relative errors in computed hyperfine
splittings of 0.6~\% and 1.4~\% for the ground and $\nu_2$ states, respectively. For other
fundamental and overtone bands,\footnote{We note that we found some inconsistencies in the
   experimental results reported in Table~VIII of reference~\onlinecite{Dietiker:JCP143:244305},
   which we have corrected in our analysis.} these errors are bigger, up to 160~kHz (3.9~\%),
however, the estimated uncertainty of the experimental data already accounts for
$\pm100$~(2.4~\%)~kHz~\cite{Dietiker:JCP143:244305}.

We believe that the accuracy of the quadrupole splittings can be significantly improved by employing
a better level of the electronic structure theory in the calculations of the EFG surface. For
example, the aug-cc-pVQZ basis set incompleteness error and the core electron correlation effects
were shown to contribute up to 0.004~a.u.\ and 0.01~a.u., respectively, into the absolute values of
the EFG tensor of the water molecule~\cite{Olsen:JCP116:1424}. By scaling these values with the
nuclear quadrupole constant of the $^{14}$N atom, we estimate that the electronic structure errors
in the quadrupole splittings of ammonia are as large as 50~kHz.

\begin{figure}
   \includegraphics[width=\linewidth]{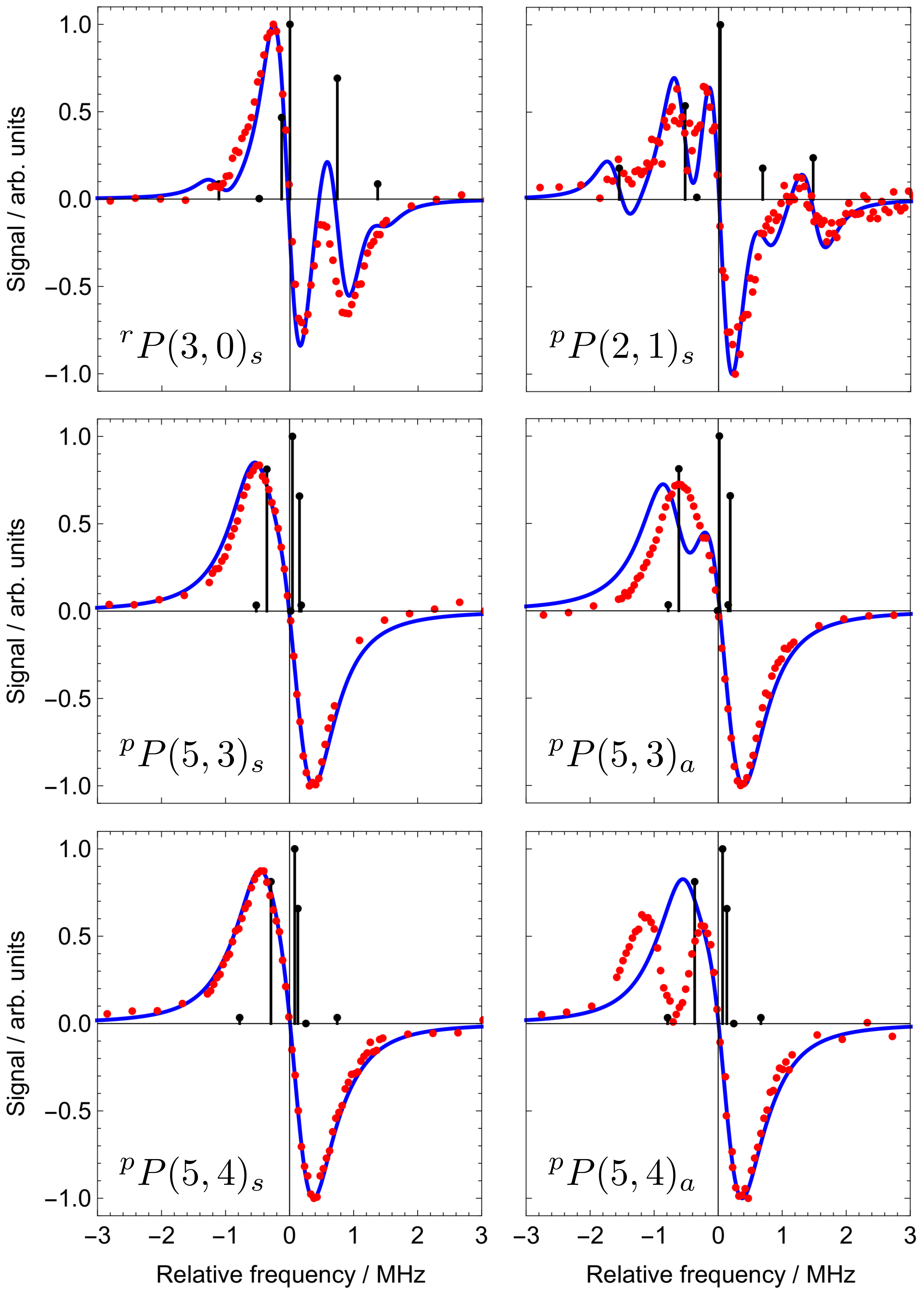}
   \caption{Comparison of the calculated (blue line) and observed \cite{Twagirayezu:JCP145:144302}
      (red dots) saturation dip line shapes for
      $^{\Delta K_a}\Delta J(J'',K_a'')_{\tau_{\rm inv}''}$ transitions of the $\nu_1+\nu_3$ band of
      NH$_3$ ($\tau_{\rm inv}''=s~{\rm or}~a$ denotes \emph{symmetric} or \emph{anti-symmetric}
      inversion parity of the ground vibrational state). Stems show contributing to the calculated
      line shape stick spectrum. The experimental and calculated intensities are normalized to the
      respective maximal values. The measured (calculated) zero-crossing wavenumbers,
      in \invcm, are 6544.32154 (6544.49333) for $^rP(3,0)_s$, 6572.85349 (6572.87926) for $^pP(2,1)_s$,
      6529.18969 (6530.047789) for $^pP(5,3)_s$,
      6528.76857 (6528.51408) for $^pP(5,3)_a$, 6536.59280 (6537.31784) for $^pP(5,4)_s$,
      and 6537.68063 (6536.68512) for $^pP(5,4)_a$.}
   \label{fig:2}
\end{figure}
In \autoref{fig:2} we compare our results with sub-Doppler saturation-dip spectroscopic measurements
for the $\nu_1+\nu_3$ band of \NHHH~\cite{Twagirayezu:JCP145:144302}. The saturation-dip lineshapes
were calculated as the intensity-weighted sums of Lorentzian-lineshape
derivatives~\cite{Axner:JQSRT68:299} with a half-width-at-half-maximum (HWHM) width of the
absorption profile of 290~kHz and the HWHM-amplitude of the experimentally applied
frequency-modulation dither of 150~kHz~\cite{Sears:privcomm:2017}. The $^pP(5,K_a'')$ transitions were recorded with
slightly larger HWHM \cite{Sears:privcomm:2017} and we found a value of 500~kHz to best reproduce
the measured lineshapes for these transitions. The computed profiles for $^rP(3,0)_s$, $^pP(2,1)_s$,
$^pP(5,3)_s$, and $^pP(5,4)_s$ transitions show very good agreement with the measurement. 
For $^pP(5,3)_a$ and $^pP(5,4)_a$ transitions the
calculated profiles do not match the experiment very well. In the experimental
work~\cite{Twagirayezu:JCP145:144302}, the predicted double peak feature of the $^pP(5,3)_a$ was not
observed while for the $^pP(5,4)_a$ it was attributed to perturbations. Based on the results of the
present variational calculations the latter can not be confirmed. It should be noted, however, that
the accuracy of the underlying PES is not sufficiently high in this energy region at 1.5~\um to
unambiguously match the predicted rovibrational frequencies with the measured ones. Moreover, the
PES employed here was obtained by a refinement of the \emph{ab initio} surface to the high
resolution spectroscopic data of NH$_3$. It is well known, that the PES refinement may cause
appearance of the spurious intensity borrowing effects as well as dissipation of the true accidental
resonances in various regions of the spectrum \cite{Yachmenev:JCP139:204308,
   AlRefaie:MNRAS448:1704}. Therefore, we refrain here from discussion of the possible alternative
assignment of the $^pP(5,3)_a$ and $^pP(5,4)_a$ transitions. Calculations of a new, more accurate
PES of NH$_3$ are currently performed and analyzed \cite{Coles:inprep}, which, once available, will
be used to generate a more accurate quadrupole-hyperfine-resolved spectrum.

In conclusion, we have presented the first general-molecule variational implementation of
nuclear-quadrupole hyperfine effects. Our approach is based on TROVE, which provides accurate
spin-free rovibrational energy levels and wave functions used as a basis for the quadrupole-coupling
problem. The initial results for \nNHHH are in very good agreement with the available experimental
data. The generated rovibrational line list for \nNHHH with quadrupole-coupling components is
available as part of the supplementary material \cite{Yachmenev:nh3_nucquad_suppinfo}. We believe that computed hyperfine-resolved
rovibrational spectrum of ammonia will be beneficial for the assignment of high resolution
measurements in the near-infrared.

Calculations based on more accurate PES and the extension of the present approach to incorporate the
hyperfine effects due to the spin-spin and spin-rotation couplings are currently performed in our
group. Due to the general approach, predictions of similar quality will be possible for other small
polyatomic molecules in order to guide future laboratory and astronomical observations with
sub-Doppler resolution, including investigations of \emph{para-ortho}
transitions~\cite{Miani:JCP120:2732, Horke:ACIE53:11965} or proton-to-electron-mass
variations~\cite{Veldhoven:EPJD31:337, Cheng:PRL117:253201}.




We gratefully acknowledge Trevor Sears for providing us with their original experimental
data~\cite{Twagirayezu:JCP145:144302}. Besides DESY, this work has been supported by the excellence
cluster ``The Hamburg Center for Ultrafast Imaging—Structure, Dynamics and Control of Matter at the
Atomic Scale'' of the Deutsche Forschungsgemeinschaft (CUI, DFG-EXC1074).

%

\end{document}